\documentclass[10pt,a4paper]{article}

\usepackage[margin=1in]{geometry}
\usepackage{times}  
\usepackage{graphicx}
\usepackage{url}
\usepackage{authblk} 
\usepackage{setspace}
\usepackage{caption}
\usepackage{subcaption}
\usepackage{amsmath, amssymb, amsthm}
\usepackage{physics}
\usepackage{xcolor}

\usepackage{hyperref}
\usepackage{float}

\title{{The Heat Kernel Expansion: Curvature for Shock Detection in Higher-Order Financial Networks}}

\author[1.2]{Mohammad Elsayed}
\author[1,2]{Sara Najem}
\affil[1]{Department of Physics, American University of Beirut, Lebanon}
\affil[2]{Complexity and Network Science Cluster at the Center for Advanced Mathematical Sciences, American University of Beirut, Lebanon}

\date{} 

\begin{document}
\maketitle

\begin{abstract}
This work follows the evolution of financial networks in Norway over a period of nine years at a monthly rate. 
The data consist of board directors and their affiliations to companies, which we model as simplicial complexes. In this framework, directors are represented  as nodes and companies as faces of the complex. 
To characterize the latter, we focus on three topological measures: the Euler characteristic, computed through the Betti numbers, torsion computed through the reduced determinant of the higher-order Laplacians, and higher-order clustering coefficients. The first two fail to capture the effect of imposed law on representation, unlike our notion of curvature which is a geometrical measure computed from the coefficients of the series expansion of the heat kernel in powers of time, which is our major contribution in this work.  In particular, the Euler characteristic integrates curvature, and thus local information is lost. Subsequently, not every topological measure can reliably capture shocks in networks. Further, the number of spanning trees may undergo significant changes at the lowest order, yet these changes need not be reflected in the torsion. Conversely, the change in the curvature revealed variation in the board interlock due to legislation, and serves as a sensitive measure for detecting shocks in networks. Inflection points in curvature are associated with external forcing, and minima with shock arrival times. Sharp transitions are also observed in the components of torsion, while smooth changes are observed in higher-order clustering. 
\end{abstract}

\vspace{1em}
    \vspace{-0.2cm}

\section*{Introduction}

Networks have been proposed as a natural mathematical representation that is capable of describing dyadic interactions and dynamics of complex systems \cite{barabasi2016network_ch1s1_5,newman2018networks}. In this framework, the individual entities are represented by nodes, and their associated interactions by links or edges \cite{barabasi2016network}, and the resulting networks can be topologically characterized, using degree distribution, centrality, clustering, and many other measures.

However, the representation of dyadic interaction falls short of capturing higher-order interactions that arise frequently in natural, physical, and social systems \cite{Bianconi_2021}. Examples include collaboration and co‑authorship networks, musical networks, neuronal networks, and interconnected infrastructure, to name just a few \cite{bardoscia2021physics,mrad2025higherordernetworkrepresentationj}. Consequently, there has been a paradigm shift toward higher-order representations using hypergraphs or simplicial complexes to model these convoluted interactions. For simplicial complexes, tools from algebraic topology and differential geometry are brought to bear, making them a powerful framework to model and probe the structure and dynamics of complex systems characterized by higher-order interactions \cite{battiston2021physics}. 

Contrary to simple graphs where a network is made up of nodes and edges, capturing only pairwise interactions,  a simplicial complex encodes the multi-node interactions making up higher-order edges starting with edges (1-simplices), triangles (2-simplices), tetrahedra (3-simplices), and more. Notions like higher-order adjacency, generalized degree distributions, clustering, and many more were introduced to study these structures \cite{Bianconi_2021}.  Key contributions also include the use of higher-order Laplacians \cite{nurisso2024higherorderlaplacianrenormalization}, which generalize graph Laplacians and capture characterizing features of simplices of any dimension.
In particular, the Euler characteristic is a topological invariant defined as the alternating sum of Betti numbers. It can be computed from the degeneracy of the zero eigenvalues of the higher‑order Laplacians; it can capture the global topology. Its connection to geometry is given by the celebrated Gauss-Bonnet theorem, which provides a link between integrated local curvature and topology. 


In network science, the link between topology and geometry remained missing until the inception of popularity versus similarity model and exponential ensembles with hidden variables \cite{Papadopoulos_2012,saucan2018discrete}, where latent geometry is  proposed as an explanation for the observed topological measures.  The characteristics of this latent geometry, combined with topology and information theory, have also been used as indicators of systemic risk \cite{keller2021hyperbolic,sandhu2016ricci,cimini2015systemic,squartini2015stationarity,samal2021network}, vulnerability \cite{gao2019measuring}, fragility \cite{barros2021using,tannenbaum2015graph,chinazzi2015systemic} communicability \cite{estrada2025forman,weber2016forman}, robustness and vulnerability, changes in flow \cite{weber2017curvature}, shocks \cite{lillo2025modeling} and more. Parallel to these geometric early warning indicators, a recent work proposed detecting structural changes in financial and economic networks using spectral methods \cite{macchiati2025spectral}, where the authors monitor the trace of the matrix exponential $\operatorname{Tr}(e^{{A}})$, where ${A}$ is the adjacency matrix of a network.
This quantity is a sum over {closed walks of any length} in the network, which are walks that start and end at the same node, thereby it encodes information about cycles, and potential pathways for shock propagation. By comparing the empirical spectral radius $\lambda_1$, or equivalently the trace, to its expected value under a null model,  they detect out-of-equilibrium behavior and systemic risk in interbank networks and international trade networks across the 2008 financial crisis.

Here we ask the question of the connection of topology to geometry using a recently introduced notion of curvature that relies on the {heat-kernel expansion}, which is a combination of geometric and spectral measures, to the context of risk and shock detection. The heat kernel describes diffusion on a network over a time scale $t$, and encodes geometric information through an asymptotic expansion. The coefficients of the expansion are geometric invariants such as scalar curvature, the Riemann and Ricci tensors. 

There is a deep mathematical connection between our approach and that of \cite{macchiati2025spectral}. The heat kernel also measures walks but, unlike the trace, it assigns a set of invariants to each node, which is information that the trace blurs. 
Furthermore, the spectral method in \cite{macchiati2025spectral}, is limited to {pairwise interactions} encoded in the adjacency matrix ${A}$. It cannot directly account for higher-order interactions (triangles, tetrahedra) that occur naturally in financial networks such as board membership data, co-authorship networks, where dynamics can propagate through groups rather than just dyadic links.
By lifting the analysis to simplicial complexes and employing the {higher-order heat kernel} defined via the higher-order Laplacian  (which acts on $k$-simplices), we generalize the spectral approach to capture walks on simplices of any dimension. 

Thus, while the spectral radius method detects global topological changes via closed walks on graphs, our curvature-based method detects {local geometric changes} arising from higher-order interactions. In the context of financial networks, specifically our case study of Norwegian board membership data before and after the 2008 gender quota reform,  we hypothesize that curvature will reveal subtle structural transitions. The imposed legislation, as an external shock, may have reorganized not just who is connected to whom, but the very geometry of how directors co-serve on boards. Tracking curvature over time provides a lens into that higher-order reorganization.
Our main hypothesis is that {curvature is the right measure for structural change detection} in higher-order systems, complementing and extending spectral methods originally developed for pairwise networks. The heat kernel expansion gives us a principled, data-driven way to compute this curvature directly from simplicial complexes without requiring an explicit metric analogous to how the spectral method works directly from adjacency matrices.

The remainder of this paper is organized as follows. Section II introduces the theoretical background on simplicial complexes and higher-order Laplacians. Section III describes our case study data. Section IV reports our empirical results. Section V concludes with a discussion of implications for systemic risk monitoring and higher-order network resilience.

\section{Theoretical Background}
\subsection{Hodge Laplacian: Geometry and Topology}
The simplicial complexes representation of networked systems is one that enables  capturing higher-order interactions present among their constituents. An $n\text{-}$dimensional simplex $\alpha_n$ is  a set of $n+1$ nodes. It is associated with an orientation $[v_0, v_1, \ldots, v_{n}] 
= (-1)^{\sigma(\pi)} 
[v_{\pi(0)}, v_{\pi(1)}, \ldots, v_{\pi(n)}]$. 
Further, the $n$-chain group $\mathcal{C}_n$ is the vector space generated by all oriented $n$-simplices of the complex.\\

\noindent The boundary map $\partial_m$ is a linear operator: 
\[
\partial_n : \mathcal{C}_n \to \mathcal{C}_{n-1},
\]
whose action is determined by: 
\[
\partial_n [v_0, v_1, \ldots, v_{n}]
= \sum_{p=0}^{n} (-1)^p \, [v_0, v_1, \ldots, \widehat{v}_p, \ldots, v_{n}],
\]
where $\widehat{v}_p$ denotes the omission of the vertex $v_p$.
To make this less abstract, we give the example of the action of the boundary operator on an edge, and on a triangle (higher-order edge). For an oriented edge $[r, s]$:
\[
\partial_1 [r, s] = [s] - [r].
\]
For an oriented triangle:
\[
\partial_2 [r, s, q] = [s, q] - [r, q] + [r, s].
\]
\vspace{0.5em}

The transpose of the matrix encapsulating the action of the boundary operator on all chains is the incidence matrix $B_{[n]}$, which are the oriented $n$ to $n+1$ connectivity operators \cite{Bianconi_2021}. This allows to build operators and compute a series of topological invariants to characterize the simplicial complex.  
Most importantly, the higher-order Laplacians can be constructed from $B_{[n]}$ through the following dependence:
	$L_{[n]} = B_{[n]}^T B_{[n]} + B_{[n+1]} B_{[n+1]}^T $. 
This also allows the construction of the block diagonal matrix of the Hodge Laplacians of all orders $L_{[n]}$ called the super Laplacian $L_s$ is given by:
\begin{equation}\label{lsup}
L_s = \begin{pmatrix}
L_{[0]} & 0 &  0\\
0 & L_{[1]} & 0\\
... & ... & ... & \\
\end{pmatrix}
\end{equation}

The spectral properties of $L_{[n]}$ reveal features of the simplicial complex in question. Particularly, the eigenvalue number density is given by a power-law dependence on the eigenvalues  $\mathcal{N}_{[n]} (\lambda' \leq \lambda)=a\lambda^{d_s^{[n]}}+b\lambda^{d_s^{[n-1]}} $, where the exponent is the spectral dimension of the $n\text{-}$order Laplacian \cite{torres2020simplicial}.

The Laplacian being a second derivative operator, is known to have ties to geometry, and curvature in particular. Specifically, geometry can be probed using the heat equation and its associated kernel,  which is encoded in the diffusion patterns over the structure given by:
  $K(t) =  e^{-t L_s}.$ 
Essentially, geometric invariants are associated with nodes, edges, faces,  and higher-order edges of arbitrary order $n$. More explicitly, due to the block structure of the Laplacian, the expansion of $K_n(i,i,t)$ close to $t\rightarrow0^+$, known as the heat-kernel expansion,  the diagonal elements of $K_n(i,i,t)$ are related to curvature through the following relation, given in \cite{najem2026geometric},  which was inspired by works on curvature on manifolds:
\begin{equation}\label{kn}
K_n(i,i,t)=
(4\pi t)^{-d_s^{[n]}/2} \sum_{k=0}^{\text{degree i}} u_{k,n}(i)\, t^k
\end{equation}
where $n$ is a face of order $n \in [0,n_{max}]$, $d_s^{[n]}$ is the spectral dimension resulting from the scaling of the cumulative number density $\mathcal{N}_{[n]}(\lambda)$ with $\lambda$, the eigenvalues of the corresponding $L_{[n]}$. The curvature $R[i]$ of the $i\text{-}$th simplex of order $n$ for example are given by: 
$u_{1,n}[i]= \frac{1}{6} R[i] - \mathcal{E},$
where $ \mathcal{E}$ is the endomorphism and for $n=0,1,2$ it is given by:
$0,R_i{}^{j}$, and $R_i{}^k + R_j{}^k - R_{ij}{}^{kl}$ respectively. Thus the coefficients correspond to scalar, mixed Ricci, and mixed Riemann curvatures. 

So far, we have presented a usage of the Laplacian as an operator which provides two geometrical invariants, however it also can reveal numerous other topological invariants. 
 For example, the degeneracy of the zero eigenvalue of $n\text{-}$th order Laplacian $L_{[n]}$ gives the Betti numbers $\beta_n$ which correspond, for each $n \geq 0$, to the number of linearly independent $n$-dimensional cavities or holes, of the simplicial complex. Their alternating sum gives the Euler characteristic given by: 
 \begin{equation}\label{eulerchar}
     \chi = \sum (-1)^n \beta_n
 \end{equation}

Additionally, it allows the computation of torsion on simplicial complexes, which is a multiscale quantity denoted by $\tau_R$, known as the Reidemeister torsion ~\cite{catanzaro2015kirchhoff,duval2009simplicial}:
\[
\log(\tau_R) = \frac{1}{2} \sum_{i=0}^{k_{\max}} i (-1)^{i+1} \log \big( \det L_{[i]}^* \big).
\label{Torsion}
\]

Here, $L_{[i]}^*$ denotes the reduced Hodge Laplacian, obtained by removing rows and columns corresponding to the harmonic subspace (i.e., restricting to the orthogonal complement of the kernel), which effectively measures the spectral signature of the holes, giving indicators about their sizes \cite{NAJEM2026118170}.

 \begin{figure}[H] 
        \centering
        \includegraphics[width=0.8\linewidth]{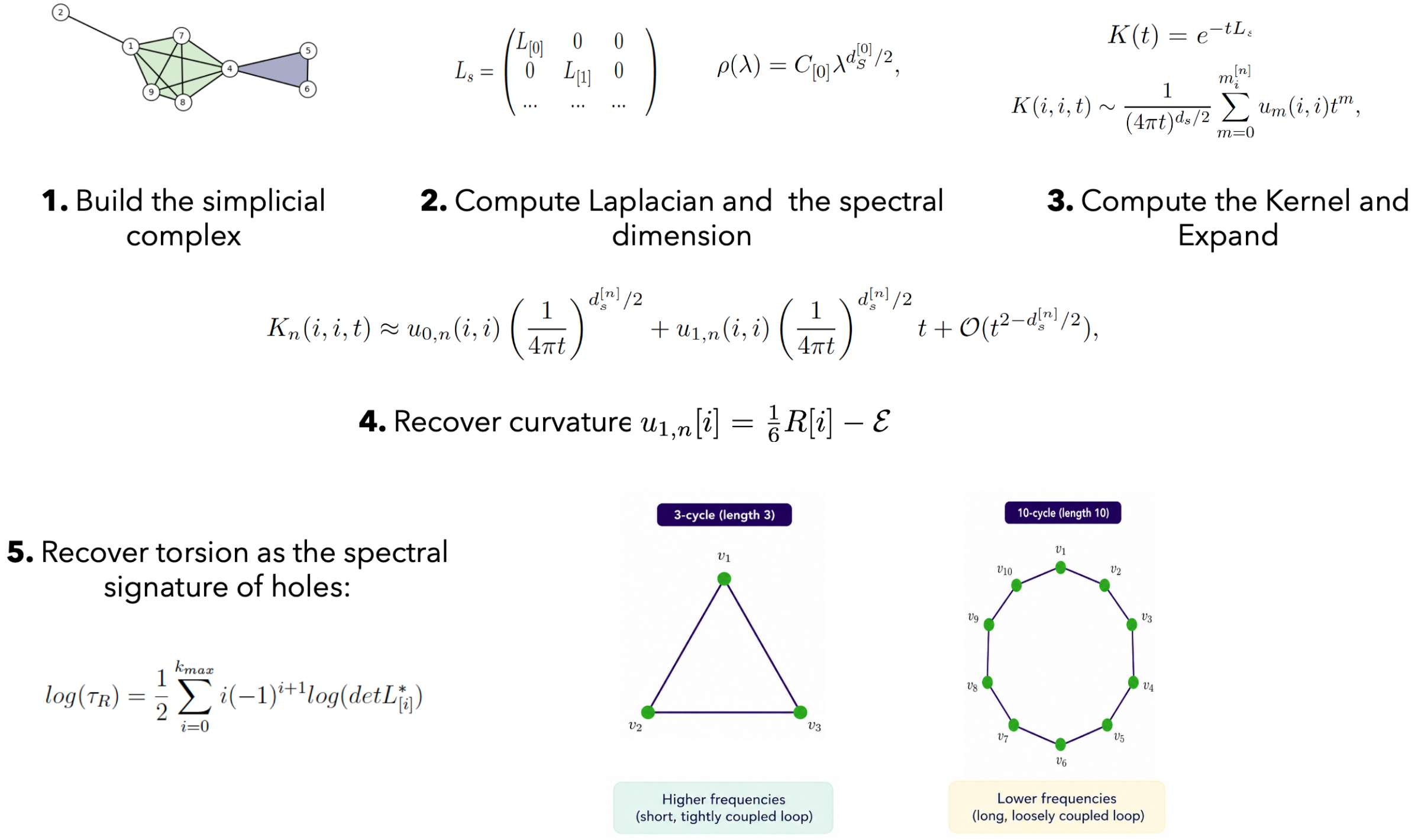}
            \caption{Work Flow Adopted in the Topological and Geometric Characterization of our Simplicial Complex}
            \label{fig:scheme}
    \end{figure}

\section{Data}
The Norwegian government, in its public limited companies act in December 2003, enforced the balance within their boards of directors \cite{seierstad2011few}. In addition to that, the government had indicated willingness to lift the law if companies were to voluntarily comply by July 2005. This attempt failed with an increase of 16 $\%$ in women representation.  Subsequently, the Norwegian government introduced in January 2006 a gender representation law that forces the composition of the board of directors of public limited companies to be at least 40$\%$ of each sex by January 2008.

The data we are using thus spans  May 2002 till August 2011 and consist of 384 public companies, which includes the People's File, or the name of directors in the companies, an associated ID and gender. It also includes the two and one-mode networks. The former involves the board-directors relations which links the ID of the director to a corresponding company, as well as the IDs of the directors making up the companies' boards. For the one-mode network, it simply encodes dyadic interactions between directors, if they sit on a common board irrespective of the latter's identity. 

We model the data and interactions between directors as a simplicial complex evolving in time. The construction of the simplicial complex goes as follows: directors are modeled as nodes and boards are the $k\text{-}$simplexes that include $k+1$ directors. A sample is shown in Figure \ref{fig:inter} below.
\begin{figure}[H] 
        \centering
        \includegraphics[width=0.3\linewidth]{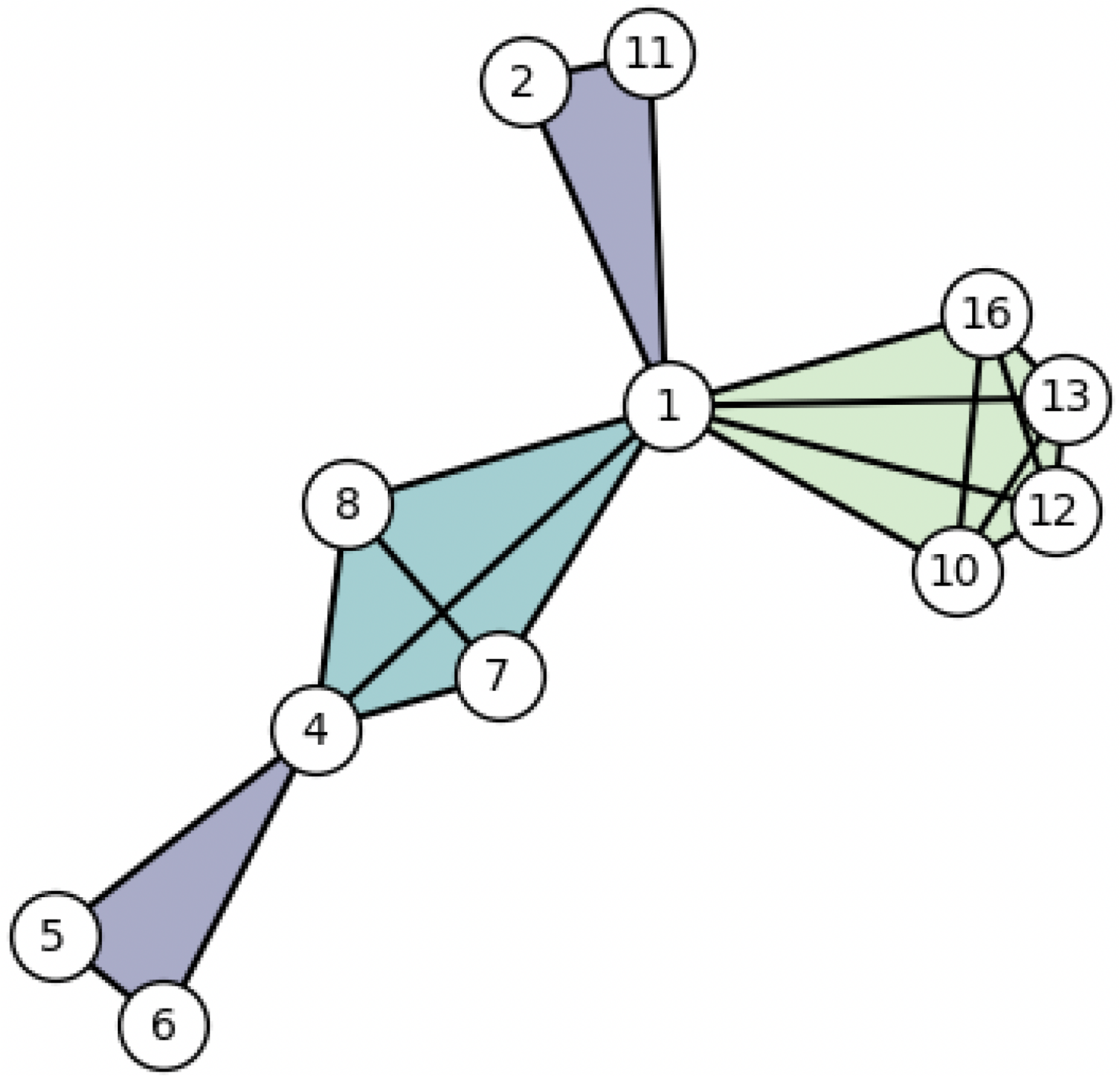}
            \caption{Sample Construction Depicting a Small Realization of Interlocking Boards}
            \label{fig:inter}
    \end{figure}
    \noindent
In figure ~\ref{fig:inter}, simplex [4,5,6] is a board that includes directors with labels 4, 5, and 6. Thus, these directors are completely connected. We can also have scenarios where there is an interlock between boards. For example, boards [1,2,11] and [1,4,7,8] are interlocking at 1. This means that director 1 sits on both boards. 
We note that with this construction, the data admits a rich encoding of higher-order interactions reaching 12 directors per board.

\section{Results and Discussion}

 The construction mentioned above is done using  {\sf ToponetX} package in {\sf python}. This package is used for building and visualizing simplicial complexes \cite{hajij2023topological}.
We first start by computing the eigenvalue decomposition for the $L_{[m]}$, and the corresponding spectral dimension for each order $d_s^{[m]}$. We then compute curvature, and follow the evolution of the total number of directors by gender over the whole period of data availability. These are shown, respectively, in Figures \ref{fig:R1evol} and \ref{fig:number} below.

 \begin{figure}[H] 
        \centering
        \includegraphics[width=0.71\linewidth]{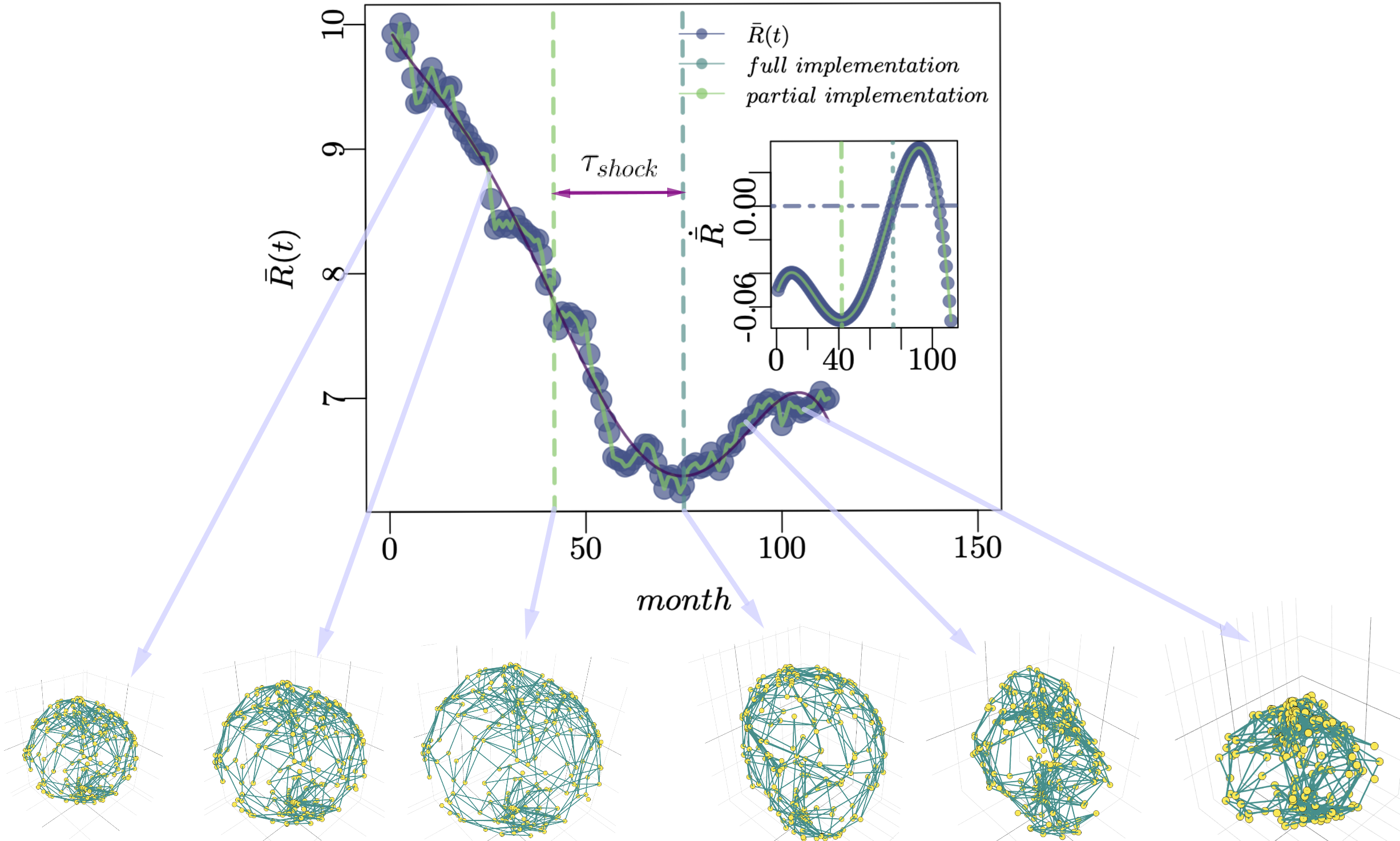}
            \caption{The average node curvature $\bar{R}^{}(t)$ is shown, fitted to a fifth-order polynomial $\dot{\bar{R}}$. $\bar{R} =  9.968 - 5.290 \times 10^{-2} t + 1.538 \times 10^{-3} t^{2} - 7.047 \times 10^{-5} t^{3} + 9.894 \times 10^{-7} t^{4} - 4.154 \times 10^{-9} t^{5}$, with R-squared is  0.98. The dashed lines correspond to the introduction of the legislation in January 2006, and its full implementation in January 2008, which coincide with the curve's inflection point and minimum.  The inset shown follows the rate of change in curvature.   }
            \label{fig:R1evol}
    \end{figure}

 \begin{figure}[H] 
        \centering
        \includegraphics[width=0.31\linewidth]{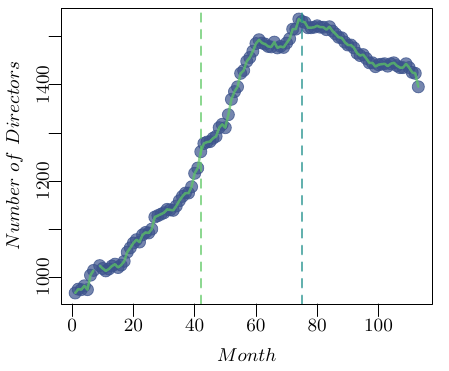}
        \includegraphics[width=0.31\linewidth]{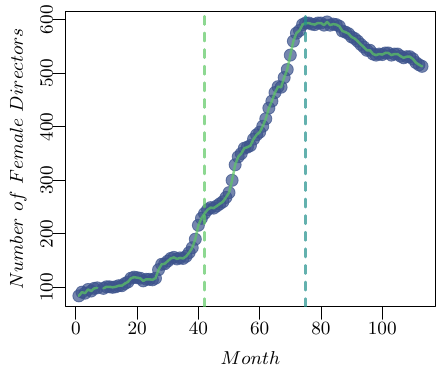}
        \includegraphics[width=0.31\linewidth]{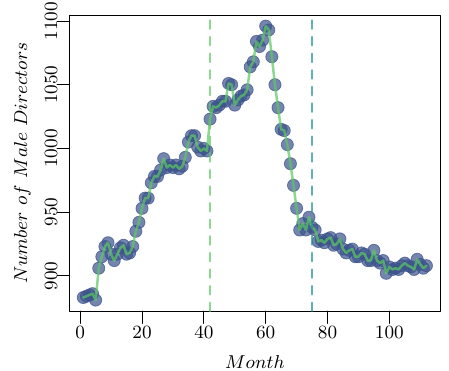}

            \caption{The figures follow the evolution of the total number of directors as well as evolution per gender. }
            \label{fig:number}
    \end{figure}
    \noindent

Specifically, using Equation \ref{kn}, we follow the monthly average edge curvature through $\bar{R}$, shown in Fig. \ref{fig:R1evol}. We note that curvature decreases to a minimum followed by a shallow increase. The data is fitted  to fifth order polynomial, which is characterized by an inflection point and a minimum, located exactly at the two critical times of legislation and the deadline for its implementation.  Thinking of the introduction of the law as an external forcing, the curvature dynamics reveal the time of propagation of the shock on companies' boards' composition, measured between the vertical dashed lines. The law was issued in January 2006 and was applicable to companies that entered into work force after that date, which is represented by the light green vertical line, which also interestingly corresponds to an inflection point, or the minimum in the rate of change of $\dot{\bar{R}}$ shown in the inset. It is also worth noting that the same behavior is captured for the node curvature $\bar{R}^{(0)}$, exhibiting the same behavior with the inflection point and minimum coinciding with the legislation time and deadline. Companies registered prior to the legislation date were given a transitional period until January 2008, represented by the second dashed line. The minimum in $\bar{R}^{(1)}$ coincides with July 2008, the point where $\dot{\bar{R}} =0$ in the inset,  beyond which $\bar{R}^{(1)}$ exhibits a rising trend as opposed to the one before the full implementation of the law. This allows us to measure a shock time $\tau_{shock} = 6$ months for the network to start exhibiting changes in trends.  

In Figure \ref{fig:R1evol}, we draw an analogy between the evolving higher-order network and a sphere whose changing radius mirrors the system's dynamics. Initially, the sphere's radius increases, causing curvature to drop. The minimum curvature corresponds, in this analogy, to the sphere at its maximum radius. At the moment of the external shock (the introduction of the gender quota law), the sphere receives an inward force that deflates it. Its radius then decreases, and curvature rises accordingly. This initial drop in curvature coincides with an increase in the number of female directors (Figure \ref{fig:number}). However, while curvature exhibits a clear inflection point precisely when the law was introduced, the evolution of director counts shows no such distinctive feature. Moreover, the timing of the shock aligns with both the curvature minimum and the peak in female director numbers.
In other words, the shock is detectable in the curvature but not in the raw count 
$N$. This out‑of‑equilibrium process, a sustained flow of female directors into boards, was accompanied by a constraint that enforced their high connectivity, increasing the demand for their representation across multiple boards simultaneously. This hypothesis can be further tested by computing the evolution of higher‑order clustering in the simplicial complex.

In simple networks, the clustering $C_{0}(i)$ is given by:
\begin{equation}
    C_{0}(i) = \frac{2E_i}{k_{i}(k_{i}-1)},
    \label{Clustering}
\end{equation}
where $k_i$ and $E_i$ are the degree of the node and the number of edges between $i$'s neighbors. The clustering coefficient generalizes to higher-order simplices. For a $k$-simplex $\sigma_i$, the $k$-th order clustering coefficient is defined as:

\begin{equation}
C_k(\sigma_i) = \frac{2 E_k(\sigma_i)}{d_k(\sigma_i)\bigl(d_k(\sigma_i)-1\bigr)},
\end{equation}
where $d_k(\sigma_i)$ is the number of $k$-simplices adjacent to $\sigma_i$, that is those that belong to a $(k+1)\text{-}$face with $\sigma_i$, and $E_k(\sigma_i)$ is the number of edges between these neighboring $k$-simplices.

\begin{figure}[H]
    \centering
    \begin{subfigure}{0.45\textwidth}
   \includegraphics[width=\linewidth]{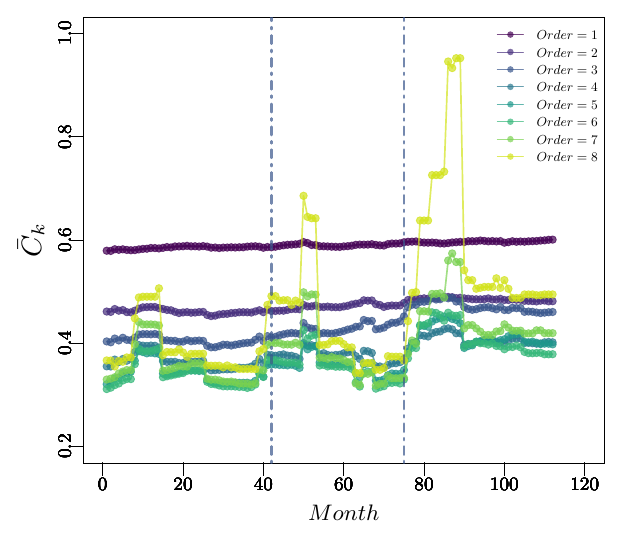}
   \caption{}
   \label{fig:Average_ClusteringDynamics}
   \end{subfigure}
      \hfill
   \begin{subfigure}{0.45\textwidth}
        \includegraphics[width=\linewidth]{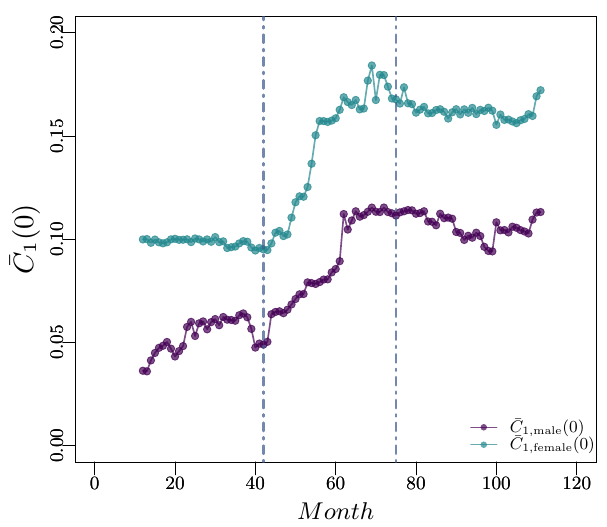}
    \caption{}
    \label{averagec1}
    \end{subfigure}
      
      \hfill

    \begin{subfigure}{0.45\textwidth}
        \centering
        \includegraphics[width=\linewidth]{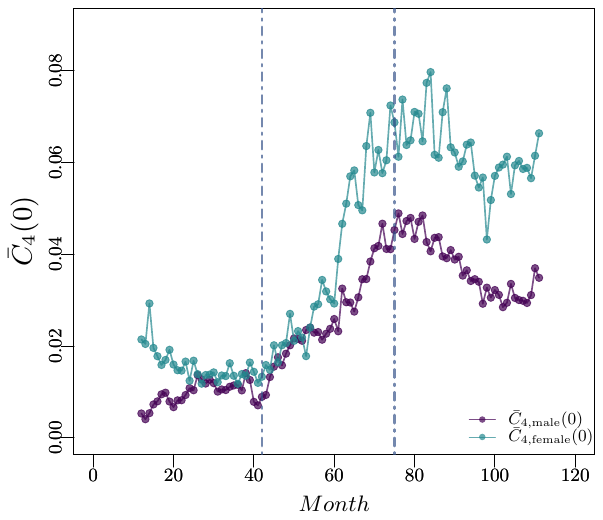}
    \caption{}
    \label{averagec4}
    \end{subfigure}
    \hfill
    \begin{subfigure}{0.45\textwidth}
        \centering
        \includegraphics[width=\linewidth]{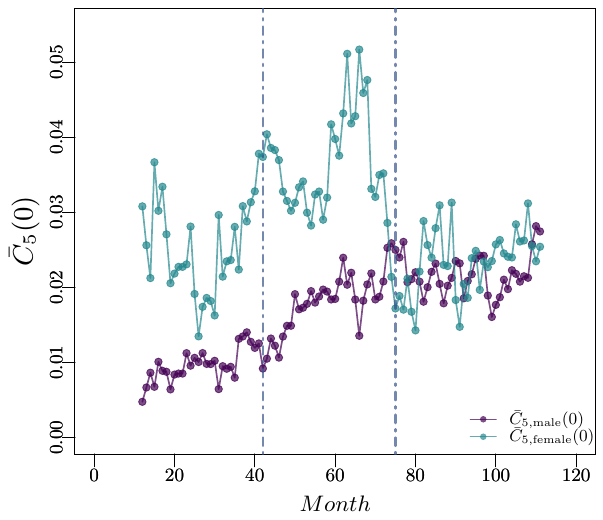}
    \caption{}
    \label{averagec5}
    \end{subfigure}
        \caption{The dynamics of the higher order clustering $\bar{C}_{k}(t)$ is shown in (a), while the evolution of female and male clustering are followed using $A_{0,m}$ for $m=1,4$ and 5 shown respectively in (b), (c), and (d). }
\end{figure}

Figure ~\ref{fig:Average_ClusteringDynamics} shows the average clustering dynamics for orders ranging from 1 to 8. We note that $\bar{C}_{1}$ and $\bar{C}_{2}$ remain constant throughout the observation period. This indicates that the interconnectedness of neighbors, whether via edges or faces is not affected by the shock. Conversely, $\bar{C}_{i\ge4}$ exhibit near periodicity in their behavior with a general increasing trend.  Almost all of them show a sharp peak right after January 2006 and a major one after January 2008. This may explain how companies managed to achieve the required percentages before the end of the deadline. For example,  $\bar{C}_{7}$ starts low, and exhibits two peaks after the vertical lines, which means that 6-simplex neighbors become increasingly connected over time and began sharing more common 7-simplices. 

\begin{figure}[H]
    \centering
    \includegraphics[width=0.5\linewidth]{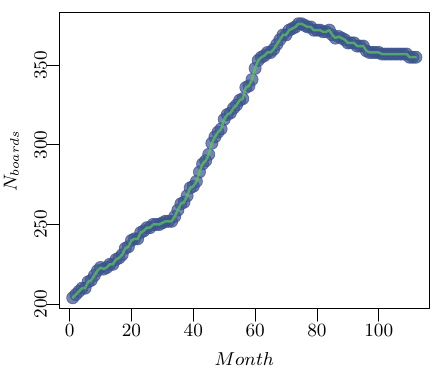}
    \caption{The figure shows the temporal evolution of the number of boards. }
    \label{fig:NbrOfboards}
\end{figure}
For further analysis, we plot the time evolution of the number of boards present in the simplicial complex, shown in Figure \ref{fig:NbrOfboards}. This quantity increases over time, peaks and then asymptotes to $\approx$ 350. The boards that join the simplicial complex interlock in a way that interconnects other boards, further increasing the interconnectedness and thus the clustering over all orders.

However, these measures blur the distinction between male and female directors, especially in the early period when the number of female directors was not comparable to that of males. To isolate the effect of the legislation on female connectivity, we focus on the adjacency matrix $A^{\times}_{(0,m)}$ and compute the node clustering separately for males and females. Figures \ref{averagec1}, \ref{averagec4}, and \ref{averagec5} show the relative difference between $\bar{C}_1(0)$ for males and females, which exhibit the most pronounced link densification, while $\bar{C}_4(0)$, and $\bar{C}_5(0)$, also show different evolutionary patterns, the effect is less pronounced than for $\bar{C}_1(0)$. 
The curves in figures ~\ref{averagec1}, and ~\ref{averagec4}, exhibit patterns expected in systems experiencing dynamical changes that force increased connectivity. The external stress on the simplicial complex is reflected in the increasing interconnectivity of both female and male directors, via edges and 4-simplices. For females, $\bar{C}_{1}(0)$ starts near 0.1, , remains constant until the law is imposed, then grows smoothly to above 0.15 before stabilizing after January 2008. For $\bar{C}_{4}(0)$ the evolution also shows growth, with female values exceeding male values, consistent with the need to increase female representation on boards.

\begin{figure}[H]
    \centering       
           \begin{subfigure}{0.45\textwidth}
        \includegraphics[width=\linewidth]{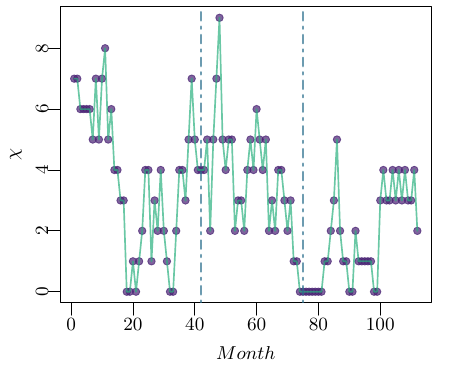}    
    \caption{}
\label{fig:chi}
    \end{subfigure}      
      \hfill
    \begin{subfigure}{0.45\textwidth}
        \includegraphics[width=\linewidth]{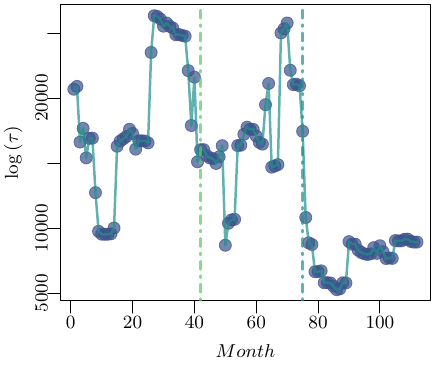}
    \caption{}
    \label{fig:torsion}
    \end{subfigure}
        \caption{The evolution of the Euler characteristic $\chi$ as well as the logarithm of torsion $\tau_R$ are shown in (a) and (b) respectively. }
        \label{fig:topo1}
\end{figure}

Now we turn to other topological measures revealed by the Laplacian. For every order $m$, which corresponds to companies with boards of $m+1$ directors, we compute the corresponding degeneracy of the zero eigenvalue of $L_{[m]}$ to obtain the Betti numbers $\beta_{[m]}$ and subsequently the Euler characteristic $\chi$. Figure~\ref{fig:chi} shows that, unlike curvature, $\chi$ exhibits no substantial change change signaling an alteration in board composition. We further compute the reduced determinant to count the number of spanning trees contributing to torsion at each order. As shown in Figure~\ref{fig:torsion}, no noticeable change is detected in $\tau_R$ either. 

\begin{figure}[H] 
        \centering
         \begin{subfigure}{0.45\textwidth}
        \includegraphics[width=\linewidth]{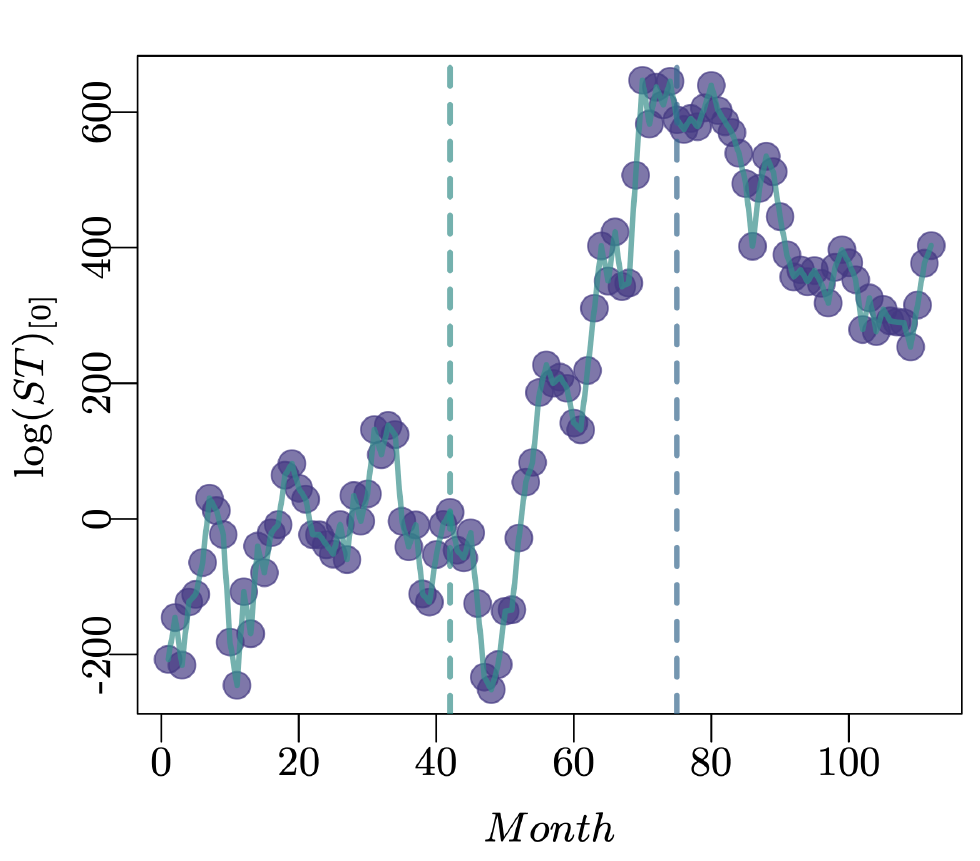}   
    \caption{}
    \label{fig:ST0}
    \end{subfigure}
        \begin{subfigure}{0.45\textwidth}
        \includegraphics[width=\linewidth]{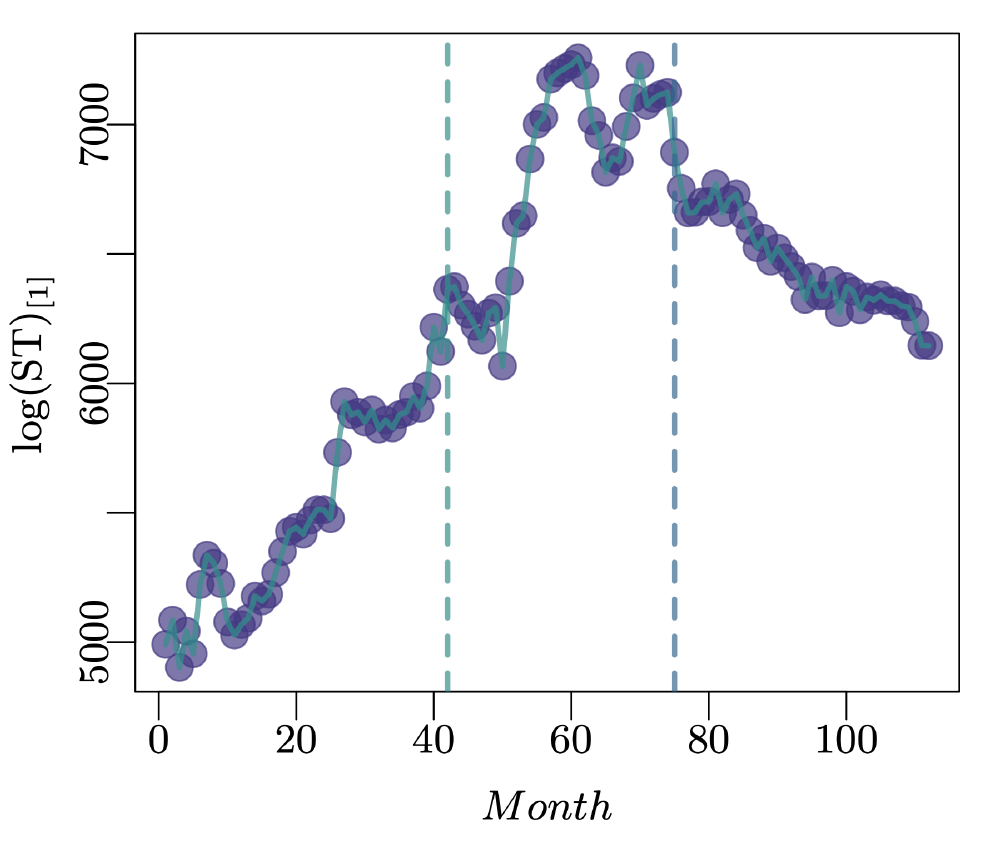}   
    \caption{}
    \label{fig:ST1}
    \end{subfigure}
        \caption{The evolution of the logarithm of spanning trees for the 0-order and first order Laplacian are shown in (a) and (b) respectively. }
        \label{fig:topo2}
    \end{figure}

To measure the effect of the higher order voids at every order, we decompose the terms contributing to $\tau_R$, and show the evolution of the 0-th order as well as the first order shown in Figures \ref{fig:ST0} and \ref{fig:ST1}. Before the law was imposed, the number of spanning trees at order zero was very low. It then grew nearly linearly from approximately 600 between January 2006 and January 2008, before dropping to around 350. This quantity peaks precisely at the imposed deadline, indicating a richer interconnected structure. However, this term has zero contribution to $\tau_R$.
To further probe the combinatorial structure, we examine the logarithm of the number of spanning trees for order 1. Although this quantity also peaks in January 2008, its near-linear growth began before the law imposition. Thus, this higher-order topological measure reveals a more sophisticated structure that would have remained hidden if we had focused only on $L_{[0]}$. The increase in the number of spanning trees is indicative of how the simplicial complex reacted to the external forcing represented by the law imposed by the government. The peaking of this quantity signals structural adaptation and reorganization of the boards of directors to achieve the 40 $\%$ female representation before the deadline. These effects at the $n\text{-}$order connectivity get blurred if we follow $\tau$, but are still captured by the growth of the spanning trees. 

\section{Conclusion}
    \vspace{-0.1cm}
In this study, we employed geometric and topological measures to characterize the structure of a simplicial complex constructed from board membership data. The geometric measure — specifically, the average curvature $\bar{R}^k$ of $k$-simplices — allowed us to identify a shock absorption time $\tau_{\text{shock}}$. This was not a quantity we had set out to measure. However, after identifying an inflection point and a subsequent minimum in curvature, we returned to the data to examine possible legislative changes. To our surprise, these structural features coincided precisely with the introduction of the gender quota law and the deadline for its implementation.

This result invites a direct parallel with recent work on detecting structural changes in financial networks using spectral methods \cite{macchiati2025spectral}. In that approach, the authors monitor the trace of the matrix exponential $\operatorname{Tr}(e^{A})$, where $A$ is the adjacency matrix of a network. This quantity sums over closed walks of any length, walks that start and end at the same node, thereby encoding information about cycles and potential pathways for shock propagation. By comparing the empirical spectral radius $\lambda_1$ (or equivalently the trace) to its expected value under a null model, they detect out-of-equilibrium behavior and systemic risk in interbank and international trade networks during the 2008 financial crisis.

Our curvature-based method shares a deep mathematical connection with this spectral approach. Both rely on spectral information: the heat kernel $(e^{-tL})$ on a simplicial complex is the natural higher-order generalization of $\operatorname{Tr}(e^{A})$ on a graph. However, while the spectral radius method aggregates information into a single scalar (the largest eigenvalue), curvature derived from the heat kernel expansion retains local geometric information and extends naturally to higher-order interactions captured by simplicial complexes. This allows us to detect not only that a shock occurred, but where and how the structure reorganized in response.

We also explored complementary topological measures and found that they did not carry the same clear signature of legislative change. The Euler characteristic, which integrates curvature, suffers from intrinsic information loss. Spanning tree counts changed significantly at the lowest order while torsion $\tau_R$, by definition, could not capture that change. These observations suggest that not all topological measures are equally capable of detecting shocks in networks. In contrast, curvature exhibited sensitivity to structural changes, a geometrically inspired measure derived from the heat kernel expansion, and proved particularly useful when the shock affects higher-order interactions that pairwise spectral methods cannot capture.

Thus, we conclude that curvature serves as a sensitive and effective observable for detecting structural changes in higher-order networks, complementing and extending spectral methods originally developed for pairwise networks. While the spectral radius detects global topological changes via closed walks on graphs, our curvature-based method detects local geometric changes arising from higher-order interactions, changes that would otherwise remain hidden.

\bibliographystyle{plainurl}
\bibliography{ref}

\end{document}